\documentclass{amsart}
\usepackage{graphicx}
\usepackage{subfigure}
\usepackage{color}
\usepackage{dsfont}

\def\mydate{\number\day\ {\ifcase\month \or January\or February\or
              March\or April\or May\or June\or July\or August\or
              September\or October\or November\or December\fi}
\number\year}

\begin{document}
\title{A Nonlocal Strain Measure for Digital Image Correlation}

\begin{abstract}
We propose a nonlocal strain measure for use with digital image correlation (DIC). Whereas the traditional notion of compatibility (strain as the derivative of the displacement field) is problematic when the displacement field varies substantially either because of measurement noise or material irregularity, the proposed measure remains robust, well-defined and invariant under rigid body motion. Moreover, when the displacement field is smooth, the classical and nonlocal strain are in agreement. We demonstrate, via several numerical examples, the potential of this new strain measure for problems with steep gradients. We also show how the nonlocal strain provides an intrinsic mechanism for filtering high frequency content from the strain profile and so has a high signal to noise ratio. This is a convenient feature considering image noise and its impact on strain calculations.
\end{abstract}

\author{R.~B.~Lehoucq}
\address{Richard B. Lehoucq, Computational Mathematics, 
  Sandia National Laboratories, P.O. Box 5800; Albuquerque, New Mexico 87185. \emph{E-mail address:} {\ttfamily rblehou@sandia.gov}, \emph{Phone: } (505) 845-8929}

\author{P.~L.~Reu}
\address{Phillip H. Reu, Sensing \& Imaging Technologies, 
  Sandia National Laboratories, P.O. Box 5800; Albuquerque, New Mexico 87185. \emph{E-mail address:} {\ttfamily plreu@sandia.gov}, \emph{Phone: } (505) 284-8913}

\author{D.~Z.~Turner}
\address{Correspondence to: Daniel Z. Turner, Multiscale Science, 
  Sandia National Laboratories, P.O. Box 5800; Albuquerque, New Mexico 87185. \emph{E-mail address:} {\ttfamily dzturne@sandia.gov}, \emph{Phone: } (505) 845-7443}

\date{\today}

\maketitle





\section{Introduction}

Digital image correlation has revolutionized material characterization by means of non-contact, full-field displacement measurement \cite{Sutton, Bruck, Hild, Chu} including several advancements towards applying DIC to characteristically difficult problems, such as extended DIC \cite{Roux, McNeill, Rethore, Rethore2, Poissant}. The reader is referred to the paper \cite{Reu1} for an overview of DIC and its applications. An important part of this process involves the mapping of displacement data with material or constitutive models via an appropriate measure of strain. There are a number of challenges to making this connection including the effects of image noise, the appropriateness of strain measures in the context of discontinuities (across pixels), and data loss from curve fitting or filtering techniques.  We propose a new nonlocal strain measure that alleviates many of these issues. 

Several strain measures have been proposed for DIC (for a review see \cite{Pan,Grama} and the reference therein). A common theme among existing strain measures involves the use of finite difference approximations of spatial partial derivatives. Although these methods have been used effectively for a variety of complex problems, there is a growing awareness of their deficiencies for problems involving cracks or strongly heterogeneous materials. Further, finite difference based approaches are sensitive to noise and often struggle to remain invariant under rigid body motion. As an alternative to these approaches, we propose a strain measure that is built on the structure of a recently proposed nonlocal vector calculus that exploits integral operators to calculate a strain.

Nonlocal vector calculus involves operators that do not use partial derivatives, but rather integrals.  A comprehensive introduction to nonlocal vector calculus is given in \cite{Du} based upon preliminary work given in \cite{GunzburgerNL}. This formalism is applied to a nonlocal diffusion equation in \cite{DuDiff} and the peridynamic Navier equation \cite{DuNL}. In the present work, we employ these ideas to construct a robust strain measure for DIC that provides many advantages over the classical strain measure. In particular, the nonlocal strain measure intrinsically filters high frequency noise from the displacement data and is well-defined even for locations at which a classical derivative may not exist.

\section{Nonlocal Strain}

We motivate the nonlocal strain measure by considering the one-dimensional problem. We then introduce the nonlocal strain measure for two and three dimensions. 
In so doing, we provide a systematic basis for computing a well-defined, discrete strain from noisy, discrete displacement data. In contrast, conventional finite difference approximations are notoriously sensitive to noisy data and inevitably lead to a tradeoff between smoothness and accuracy.

\subsection{One dimension} 

When a function $f$ is continuous at $x$, then we have the well-known relationship
\begin{subequations}
\begin{align}
\int_{-\infty}^{\infty} f(y) \, \delta(y - x) \; \text{d}y &= f(x)\,, \label{eq:fZero}
\intertext{where $\delta(x)$ is the Dirac delta (generalized) function. Probably less well-known is the relationship}
\int_{-\infty}^{\infty} f(y) \, \delta^\prime(y - x) \; \text{d}y &= -f^\prime(x)\,. \label{eq:antisymZero}
\end{align}
when $f^\prime$ is continuous at $x$.
This is easily established by integrating by parts the lefthand side to obtain
\[
\begin{aligned}
\int_{-\infty}^{\infty} f(y) \, \delta^\prime(y - x) \; \text{d}y & = f(x) \, \delta(x)\bigg|_{-\infty}^{\infty} -  \int_{-\infty}^{\infty} f^\prime(y) \, \delta(y - x) \; \text{d}y  \\
& = -f^\prime(x)\,,
\end{aligned}
\]
where we used \eqref{eq:fZero} and the formal property
\[
\delta(x) = 
\begin{cases}
0 & x \neq 0\\
\text{undefined} & x = 0
\end{cases}\,.
\]
By selecting $f=1$ in \eqref{eq:fZero} and \eqref{eq:antisymZero}, we obtain two useful relations
\begin{equation}
\int_{-\infty}^{\infty} \delta(y - x) \; \text{d}y = 1\,, \quad \int_{-\infty}^{\infty} \delta^\prime(y - x) \; \text{d}y = 0\,.
\end{equation}
In so many words, $\delta(x)$ and $\delta^\prime(x)$ are even and odd functions. More generally, the former and latter functions are symmetric and antisymmetric about the origin, respectively. 
\end{subequations}

The above discussion suggests a definition for the nonlocal derivative of $f$ at $x$ as
\begin{equation} \label{nl-deriv-1d}
-\int_{-\infty}^{\infty} f(y) \, \alpha_\epsilon(y - x) \; \text{d}y\, \text{ where } \int_{-\infty}^{\infty} \alpha_\epsilon(y - x) \; \text{d}y = 0\,.
\end{equation}
The function $\alpha_\epsilon$ is an integrable approximation to $\delta^\prime(x)$ representing the kernel of the integral operator and the parameter $\epsilon$ is a length scale associated with the approximation. This length scale can, for example, denote the nonzero region of compact support for $\alpha_\epsilon$. 

The following example clarifies the procedure used to construct an appropriate $\alpha_\epsilon$ and the role of $\epsilon$.
Consider the function 
\begin{subequations} \label{eq:HatFunction}
\begin{align}
\phi_\epsilon (x) & =  \begin{cases}    
\displaystyle 2\frac{x + \gamma \epsilon}{\gamma \, \epsilon^2} \quad  &-\gamma \,\epsilon < x < 0\\[1.5ex]
\displaystyle    2 \frac{(1-\gamma)\,\epsilon -x }{(1-\gamma)\epsilon^2}\quad  &0 < x < (1-\gamma) \,\epsilon \\
    0 & \quad \text{otherwise}
  \end{cases}
\intertext{ with derivative}
\phi^\prime_\epsilon (x) & =  \begin{cases}
     \displaystyle  \frac{2}{\gamma \, \epsilon^2}\quad &-\gamma \,\epsilon < x < 0 \\[1.5ex]
     \displaystyle    - \frac{2 }{(1-\gamma)\,\epsilon^2}\ \quad  &0 < x < (1-\gamma) \,\epsilon \\
    0 & \quad \text{otherwise}
  \end{cases}
\end{align}
where $\gamma$ is a dimensionless skew parameter satisfying the constraint
\begin{align}
0 < \gamma < 1\,. \label{gamma-constraint}
\end{align}
A simple calculation reveals that
\begin{equation}
\label{eq:SimpleCalcs}
\int_{-\gamma \,\epsilon}^{(1-\gamma)\epsilon} \phi_\epsilon(x) \, dx =1\,, \quad \int_{-\gamma \,\epsilon}^{(1-\gamma)\epsilon} \phi^\prime_\epsilon(x) \, dx = 0\quad \text{for all} \quad \epsilon>0\,,
\end{equation}
when $\gamma$  is greater than $0$ and strictly less than $1$. Otherwise, the properties \eqref{eq:SimpleCalcs} no longer hold.
Therefore  as $\epsilon \to 0$
\begin{equation}
\begin{aligned}
\int_{-\gamma \,\epsilon}^{(1-\gamma)\epsilon} f(y) \, \phi_\epsilon(x-y) \, dy & \to f(x) \\
- \int_{-\gamma \,\epsilon}^{(1-\gamma)\epsilon} f(y) \, \frac{\phi^\prime_\epsilon(x-y)}{2} \, dy & \to f^\prime(x)\,,
\end{aligned}
\end{equation}
\end{subequations}
where the limits hold if $f$ is differentiable at $x$.  We remark that if $f$ and $\phi_\epsilon$ have units of length and per length, respectively, then the nonlocal derivative is a dimensionless quantity.

\begin{figure}
  \centering
    \includegraphics[scale=0.75]{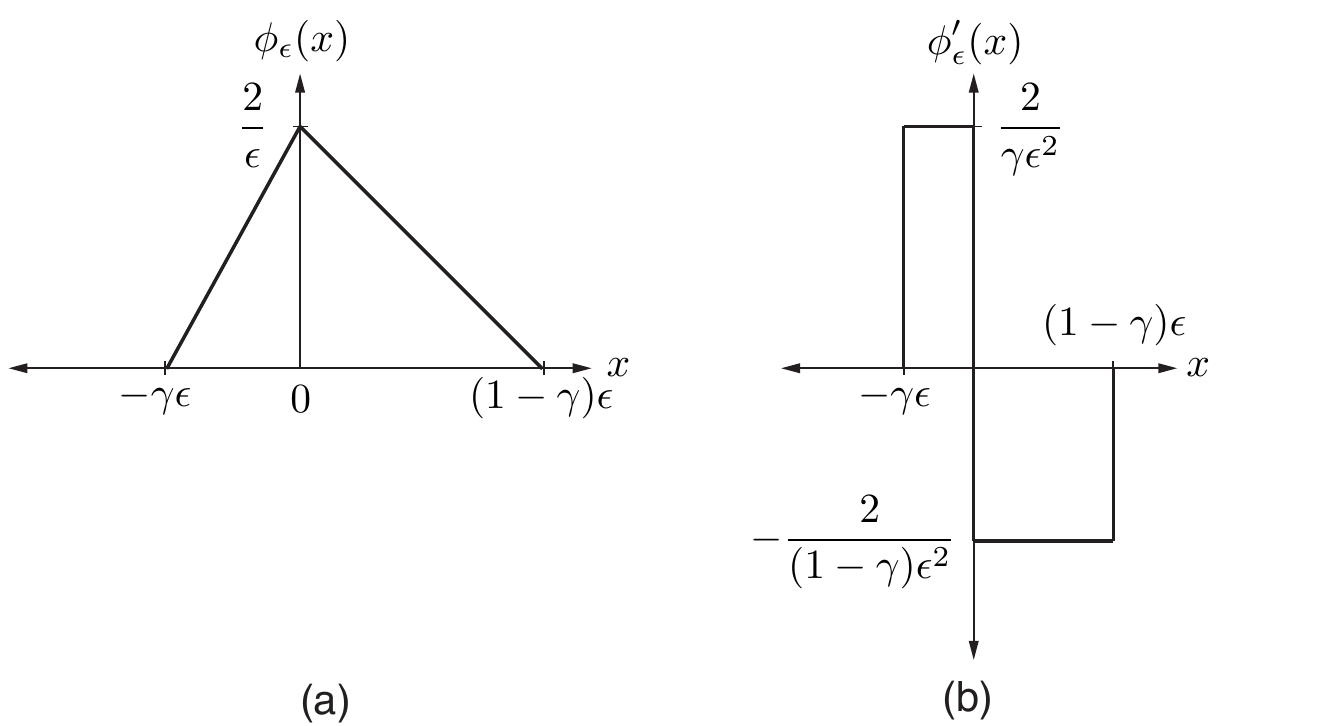}
  \caption{(a) Kernel function $\phi_\epsilon(x)$ and (b) its derivative.}
    \label{fig:Phi}
\end{figure}

Plots of this kernel and its derivative are shown in Figure \ref{fig:Phi}. The plots indicate that $\gamma$ skews the function $\phi$ into the right half of the plane. This is a useful feature when $x$ is near the end point of an interval over which approximations to $f^\prime(x)$ are needed. 

Selecting $f(x)=ax+b$ and using a midpoint quadrature rule, which is exact for linear functions, grants
\begin{equation}
\begin{aligned}
\int_{-\gamma \,\epsilon}^{(1-\gamma)\epsilon} \big( ay+b\big) \, \phi_\epsilon(x-y) \, dy &  = ax+b\,, \\
 -\int_{-\gamma \,\epsilon}^{(1-\gamma)\epsilon} \big( ay+b\big)\, \frac{\phi^\prime_\epsilon(x-y)}{2} \, dy & = a\,,
\end{aligned}
\end{equation}
so that the derivative of a linear function is computed exactly with a simple quadrature rule. Alternatively, these two relations can be established via integration by parts.

The advantage of the nonlocal derivative is that it provides a systematic and robust basis for approximating $f^\prime(x)$ given smooth \emph{or irregular data}, avoiding the characteristic sensitivity of finite difference approximations to noisy data. 

\subsection{Two and three dimensions}

Define the nonlocal gradient of a vector function $\mathbf{f}$ as
\begin{align}
\tilde{\nabla}\, \mathbf{f}(\mathbf{x}) &:=  -\int_{\mathbb{R}^n} \mathbf{f}(\mathbf{y})  \otimes\boldsymbol{\alpha}_\epsilon(\mathbf{y} - \mathbf{x}) \, \text{d}\mathbf{y} \label{nl-deriv-nd}
\intertext{where  $ \mathbf{x}\otimes\mathbf{y}$ denotes the dyadic product of $ \mathbf{x}$ and $ \mathbf{y}$,  $\boldsymbol{\alpha}_\epsilon$ is the kernel of the integral operator approximating $\nabla\delta(\mathbf{x})$ satisfying}
\mathbf{0} &= \int_{\mathbb{R}^n}  \boldsymbol{\alpha}_\epsilon(\mathbf{y} - \mathbf{x})  \, \text{d} \mathbf{y}\,, \notag
\end{align}
and $n=1,2,3$ (when $n=1$, then \eqref{nl-deriv-1d} and \eqref{nl-deriv-nd} coincide).
The parameter $\epsilon$ is a length scale associated with the approximation. This length scale can, for example, denote the nonzero region of compact support for $\boldsymbol{\alpha}_\epsilon$. The nonlocal derivative \eqref{nl-deriv-nd} is a variation of the corresponding deformation gradient tensor $\mathbf{\bar{F}}$ proposed in \cite[p.180]{Silling1} and also the weighted nonlocal gradient proposed in \cite[pp.520-527]{Du}. This weighted gradient assumed that $\boldsymbol{\alpha}_\epsilon(\mathbf{y} - \mathbf{x})+\boldsymbol{\alpha}_\epsilon(\mathbf{x} - \mathbf{y})=\mathbf{0}$, a sufficient, but not necessary, condition to satisfy the second integral in \eqref{nl-deriv-nd}. As discussed in the one-dimensional case, if $\mathbf{f}$ and $\boldsymbol{\alpha}_\epsilon$ have units of length and per volume per length, then the nonlocal gradient is a dimensionless quantity.

\begin{figure}[h!]
  \centering
    \includegraphics[scale=0.35]{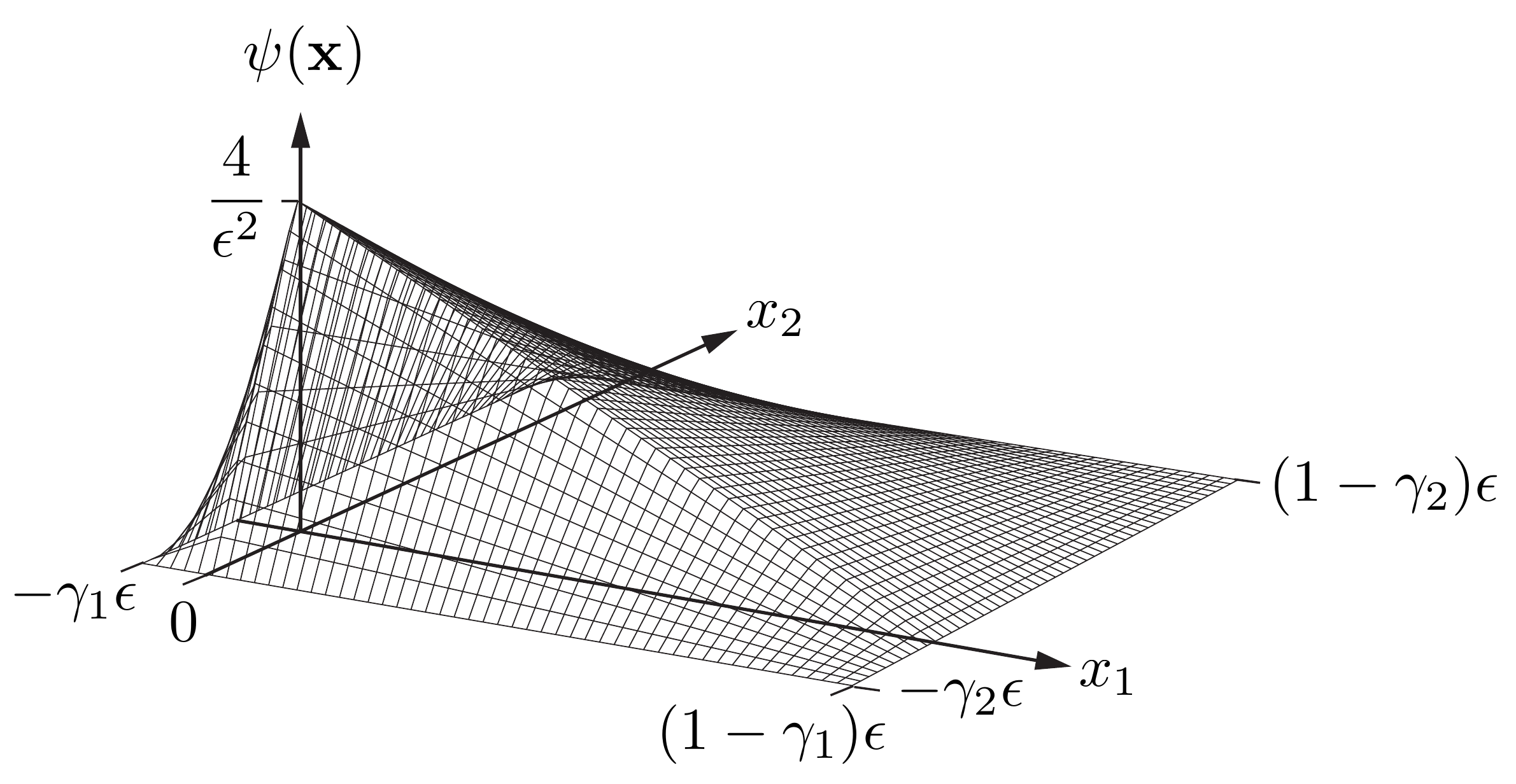}
  \caption{Plot of $\psi(\mathbf{x})$.}
    \label{fig:psi}
\end{figure}

\begin{figure}[h!]
  \centering
      \includegraphics[scale=0.35]{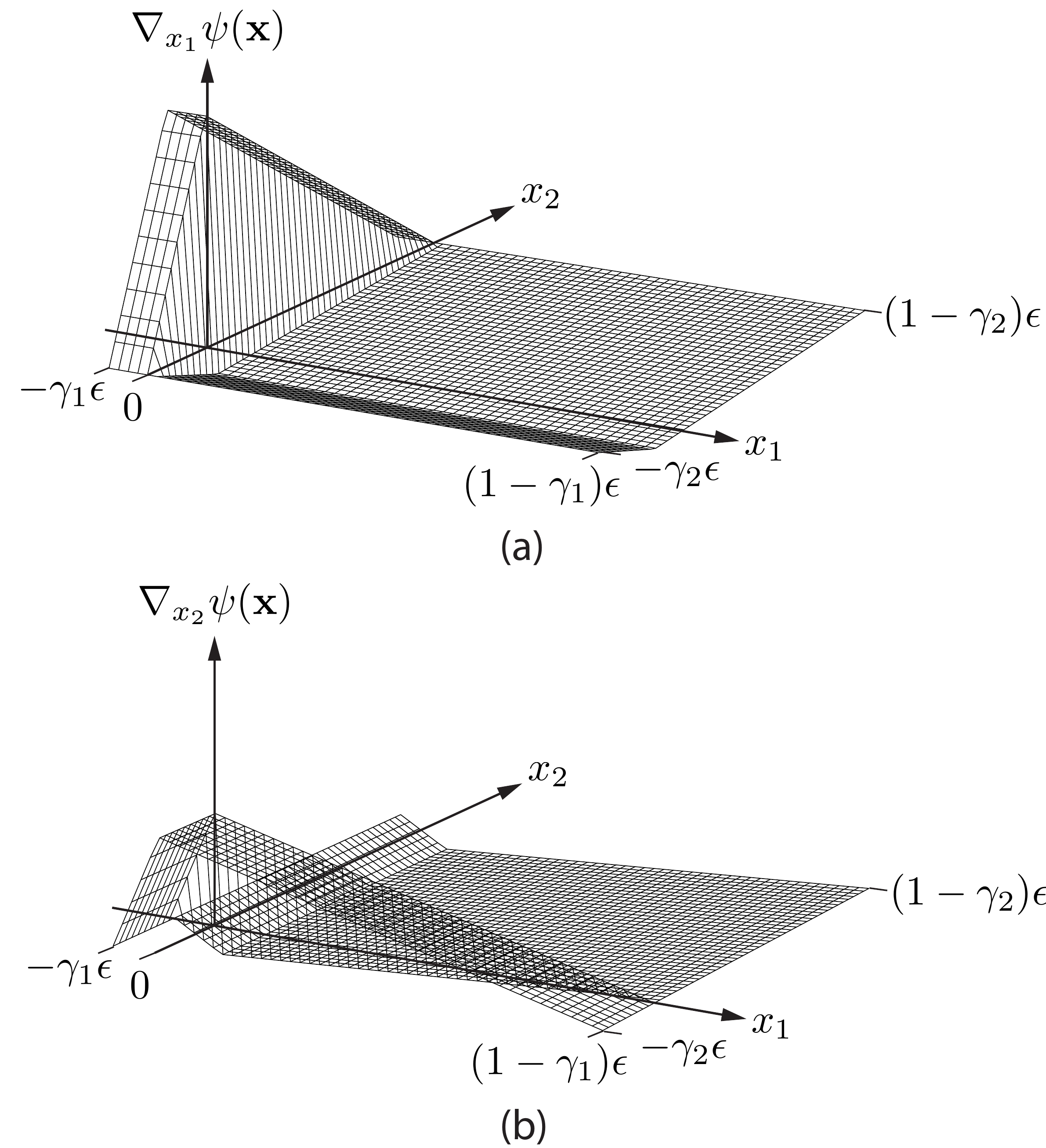}
      \caption{Plot of the components of $\boldsymbol{\alpha}_\epsilon(\mathbf{x})$, (a) $\nabla_{x_1} \psi(\mathbf{x})$ (b) $\nabla_{x_2} \psi(\mathbf{x})$}
   \label{fig:GradPsi}
\end{figure}

The determination of an appropriate $\boldsymbol{\alpha}_\epsilon$ follows as in the one-dimensional case. As an example, let the scalar function 
\[
\psi(\mathbf{x}) = \psi(x_1,\, x_2) = \phi_{1,\epsilon}(x_1)\, \phi_{2,\epsilon}(x_2)\,,
\]
where $\phi_{1,\epsilon}\,,\phi_{2,\epsilon}$ are the one-dimensional functions given by \eqref{eq:HatFunction} with skew parameters $0<\gamma_1<1$, and $0<\gamma_2<1$, respectively (see Figure~\ref{fig:psi}).  
Therefore
\[
\boldsymbol{\alpha}_\epsilon(\mathbf{x}) = \nabla \psi(\mathbf{x}) 
= \bigg(\frac{\partial \phi_{1,\epsilon}(x_1)}{\partial x_1}\phi_{2,\epsilon}(x_2) \,, \frac{\partial \phi_{2,\epsilon}(x_2)}{\partial x_2 }\phi_{1,\epsilon}(x_1) \bigg)\,.
\]
A plot of the components of $\boldsymbol{\alpha}_\epsilon(\mathbf{x})$ is shown in Figure~\ref{fig:GradPsi}.
By construction, we see that
\[
\int_{\mathbb{R}^2}  \psi(\mathbf{x})\,\text{d} \mathbf{x} = 1\,,  \text{ and } \int_{\mathbb{R}^2}  \nabla \psi(\mathbf{x}) \,\text{d} \mathbf{x} = \mathbf{0}\,.
\]

\subsection{Homogenous deformation}

Selecting the function $\mathbf{f} = \mathbf{A\,x} + \mathbf{c}$, where $\mathbf{A}$ and $\mathbf{c}$ are a constant tensor and vector, results in
\begin{equation*}
\begin{aligned}
\tilde{\nabla}\, (\mathbf{A\,x} + \mathbf{c}) &=  \tilde{\nabla}\, \mathbf{A\,x} + \tilde{\nabla}\,\mathbf{c} \\
& = -\mathbf{A}\int_{\mathbb{R}^2} \mathbf{y} \otimes\boldsymbol{\alpha}_\epsilon(\mathbf{y} - \mathbf{x}) \, \text{d}\mathbf{y} + \mathbf{c} \otimes \int_{\mathbb{R}^2} \boldsymbol{\alpha}_\epsilon(\mathbf{y} - \mathbf{x}) \, \text{d}\mathbf{y}\\
&=  -\mathbf{A}\int_{\mathbb{R}} \int_{\mathbb{R}}  (y_1,\, y_2 )   \\
&\qquad \; \otimes\bigg(\frac{\partial \phi_{1,\epsilon}(y_1-x_1)}{\partial y_1} \phi_{2,\epsilon}(y_2 - x_2) \,, \frac{\partial \phi_{2,\epsilon}(y_2-x_2)}{\partial y_2 } \phi_{1,\epsilon}(y_1 - x_1) \bigg) \\
& \qquad \quad \text{d} y_1\, \text{d} y_2\\
&= -\mathbf{A} \big(  -\mathbf{I}\big) = \mathbf{A}\,,
\end{aligned}
\end{equation*}
where $\mathbf{I}$ denotes the unit tensor and the fourth equality follows from integrating by parts each of the four components of the tensor. 


In particular, the above derivation demonstrates that
\begin{equation}\label{nl-grad-linear-trans}
\tilde{\nabla}\, (\mathbf{A\,x}) = \mathbf{A} \tilde{\nabla}\, \mathbf{x} = \mathbf{A}\,,
\end{equation}
so that $\tilde{\nabla\tilde}$ commutes with a constant tensor exactly as does the classical gradient operator. Moreover,  when $\mathbf{A}=\mathbf{I}$ then $\tilde{\nabla}\, \mathbf{x} = \mathbf{I}$, or in words, the nonlocal gradient of the identity vector map is the identity tensor.

\subsection{Definition of nonlocal strain}\label{sec:nl-str}

Consider the  nonlocal  deformation gradient defined as
\begin{subequations}
\begin{align}
\tilde{\mathbf{F}} & := \mathbf{I} + \tilde{\nabla}\mathbf{u} \,. \label{nl-def-grad}
\intertext{The nonlocal strain tensor, $\tilde{\mathbf{E}}$,  is then defined as} 
\label{eq:NonlocalStrain}
\tilde{\mathbf{E}} & := \frac{1}{2}\big( \tilde{\mathbf{F}}^T\tilde{\mathbf{F}} -\mathbf{I} \big)\,.
\end{align}
\end{subequations}

\subsection{Rigid body motions and finite rotations}

For this nonlocal strain measure to be of meaningful value, the property $\tilde{\mathbf{F}}^T\tilde{\mathbf{F}} = \mathbf{I}$ must hold for any motion that does not deform the body. In other words, the nonlocal strain should be invariant under rigid body motion defined by the mapping 
\begin{subequations}
\begin{align} 
\zeta(\mathbf{x}) &= \mathbf{R}\, \mathbf{x} + \mathbf{c} \label{urb}
\intertext{with rigid body displacement}
\mathbf{u}_{rb}(\mathbf{x}):= \zeta(\mathbf{x}) &-  \mathbf{x}  =  \big(\mathbf{R} -\mathbf{I}\big)\, \mathbf{x} + \mathbf{c} \,,
\intertext{where $\mathbf{R}$ is a rotation tensor, i.e.,  $\mathbf{R}^T \mathbf{R}=\mathbf{I}$ with positive determinant, and $\mathbf{c}$ is a constant vector. The corresponding nonlocal deformation gradient is}
\mathbf{I} + \tilde{\nabla}\mathbf{u}_{rb}  
&= \mathbf{I} +  \big(\mathbf{R}-\mathbf{I}\big) = \mathbf{R}\,,
\end{align}
\end{subequations}
where we used \eqref{nl-grad-linear-trans} for the first equality so that
\[
\big( \mathbf{I} + \tilde{\nabla}\mathbf{u}_{rb} \big)^T\big( \mathbf{I} + \tilde{\nabla}\mathbf{u}_{rb} \big) = \mathbf{I}
\]
as required for the nonlocal strain to be invariant under rigid body motion.

\subsection{Nonlocal strain over a bounded domain}

Let $\Omega$ represent a bounded open domain partitioned into $M$ non-overlapping subdomains $\Omega_i$ with boundary $\Gamma$  such that 
\begin{equation}
\begin{aligned}
  \Omega = \overset{M}{\underset{i = 1}{\bigcup}} \; \Omega_i \;.
\end{aligned}
\end{equation}
In the context of DIC, $\Omega_i$ represents the decomposition of data as either pixels or collections of pixels with an associated area. 

Let $\mathbf{u}_i$ denote the displacement over $\Omega_i$. Since the value of $\mathbf{u}_i$ is constant over $\Omega_i$, the displacement vector field over $\Omega$ is
\begin{align}\label{dispuO}
\mathbf{u}(\mathbf{x}) = \sum_{j=1}^M \mathbf{u}_i \, \mathds{1}_{\Omega_j}(\mathbf{x})
\end{align}
where the indicator function $\mathds{1}_{\Omega_j}$ is given by
\[
\mathds{1}_{\Omega_j}(\mathbf{x}) := 
\begin{cases}
1 & \mathbf{x} \in \Omega_j \,,\\
0 & \mathbf{x} \notin \Omega_j\,. 
\end{cases}
\]
Inserting the DIC displacement \eqref{dispuO} into~\eqref{nl-deriv-nd} results in
\begin{subequations} \label{nl-discrete-deriv-nd}
\begin{align}
\tilde{\nabla}\, \mathbf{u} (\mathbf{x}) & =  - \sum_{j=1}^M \mathbf{u}_j  \otimes \int_{\Omega_j} \nabla \psi(\mathbf{y} - \mathbf{x}) \, \text{d}\mathbf{y}\ \qquad \mathbf{x} \in \Omega\,,
\intertext{or equivalently,}
\tilde{\nabla}\, \mathbf{u} & = - \sum_{j=1}^M \mathbf{u}_j  \otimes  \tilde{\nabla}\,\mathds{1}_{\Omega_j} \text{ over } \Omega\,.
\end{align}
\end{subequations}
These two relations explain that the function $\nabla \psi(\mathbf{x})$ need not be evaluated directly, but rather only its integral over the subdomain $\Omega_j$ is needed. This is an important result, given that $\nabla \psi(\mathbf{x})$ is not defined at the origin or along the axes.

A special case occurs when $\mathbf{c} = \mathbf{u}_1=\mathbf{u}_2=\cdots=\mathbf{u}_M$ so that $\mathbf{u}(\mathbf{x}) = \mathbf{c} \, \mathds{1}_{\Omega}(\mathbf{x})$ and
\begin{align*}
\tilde{\nabla}\, (\mathbf{c} \, \mathds{1}_{\Omega}) & = \mathbf{c} \, \tilde{\nabla}\, \mathds{1}_{\Omega}\,.
\intertext{If we suppose that $\Omega$ is a region such that for any $\mathbf{x} \in \Omega$, there exists skew parameters $\gamma_1$ and $\gamma_2$ that satisfy the constraint \eqref{gamma-constraint} (the support of $\psi(\mathbf{x})$ is contained within $\Omega$) then}  
\tilde{\nabla}\, \mathds{1}_{\Omega}(\mathbf{x})  & \equiv \mathbf{0} \qquad \mathbf{x} \in \Omega\,.
\end{align*}
For example, if $\Omega$ is a rectangle, then the above conditions can be fulfilled since either $\gamma_1$ or $\gamma_2$ (or both) can be selected appropriately when $\mathbf{x} $ is close to the boundary, $\Gamma$, of $\Omega$. 

\subsection{Discrete approximation}

The invariance under rigid body motion for the discrete approximation to $\tilde{\mathbf{E}}$ also holds assuming a sufficiently accurate quadrature rule is used to integrate
$ \nabla \psi$ and $\mathbf{u}$. 
This can be easily verified. Consider a square, two-dimensional domain of size $L \times L$, where $L = 100$ units, with a synthetic rigid body displacement field of the form
\begin{equation}
\mathbf{R} = \left[ \begin{array}{cc}
\text{cos}\;\theta & -\text{sin}\;\theta \\
\text{sin}\;\theta & \text{cos}\;\theta  \end{array} \right] \; , \qquad \mathbf{c} = \mathbf{0} \; ,
\end{equation}
where $\theta = 45$ degrees. The domain is discretized into cells (pixels) of size one unit $\times$ one unit. Computing the nonlocal strain according to \eqref{eq:NonlocalStrain} for various values of $\epsilon$ results in the strain values given in Table~\ref{tab:RBMError}, which shows that the strain is indeed negligible.
\begin{table}[htb]
  \caption{$L_2$-norms of the nonlocal strain for various support sizes\label{tab:RBMError}}
  \centering
  \begin{tabular}{l l l l l}
    \hline
    $\epsilon$ & $\tilde{\mathbf{E}}_{11}$ & $\tilde{\mathbf{E}}_{12}$ & $\tilde{\mathbf{E}}_{21}$ & $\tilde{\mathbf{E}}_{22}$   \\
    \hline
10& 5.9e-15& 6.9e-15& 6.9e-15& 1.1e-14\\
8& 5.3e-15& 6.7e-15& 6.7e-15& 1.2e-14\\
6& 7.0e-15& 9.1e-15& 9.1e-15& 1.7e-14\\
4& 9.9e-15& 1.0e-14& 1.0e-14& 1.8e-14\\
2& 2.0e-14& 2.0e-14& 2.0e-14& 2.7e-14\\
    \hline
  \end{tabular}
\end{table}

As an aside, when the displacement is given by \eqref{dispuO}, care needs to be taken to align the discontinuity of $\nabla \psi$ over the pixel containing the origin with a pixel boundary for the quadrature involving $\nabla \psi$ since by \eqref{nl-discrete-deriv-nd} the displacement is constant over $\Omega_j$. We also remark that the nonlocal strain at a point depends upon all the values of the displacement field  surrounding the point. The quadrature for the discrete approximation of the nonlocal strain is akin to using an extremely high order differencing scheme, except that the stencil is over the full area of the nonlocal support, not just along the coordinate directions aligned with the positive and negative $x_1$ and $x_2$ axes (as it is in finite difference difference approximations to the classical gradient).

\subsection{Virtual strain guage}

For the purpose of comparison, we include here a brief description of how strain is typically calculated in the context of DIC. There are a number of nuances related to the various methods available, but in general, they fall into two categories: those that apply finite differencing to the displacement data and those that that use polynomial smoothing. In either case, filtering may also be applied. In the case of local polynomial smoothing, a polynomial function is fit to the displacement data over a particular region. This region is defined by the strain window size, which can be considered as a regularization parameter for a given polynomial basis. It can be shown that this process effectively acts as a low pass filter. The resulting strain measure is highly sensitive to the virtual strain gauge (VSG) size  (and the subset size), where the VSG size is a function of the strain window and step size. For small VSG sizes, the resulting strain field will be more accurate, but contain a large amount of noise. Conversely, a large VSG size limits the amount of noise, but tends towards less accuracy. In the numerical examples below, we perform a comparison between the VSG approach and the nonlocal strain measure to illustrate the differences between the two.

\section{Numerical Results}

In this section we demonstrate the performance of the nonlocal strain for a number of examples including verification problems and results for images taken from experiments.

\subsection{Non-fully-differentiable function on a bounded domain}

\begin{figure}[h!]
  \centering
    \includegraphics[scale=0.8]{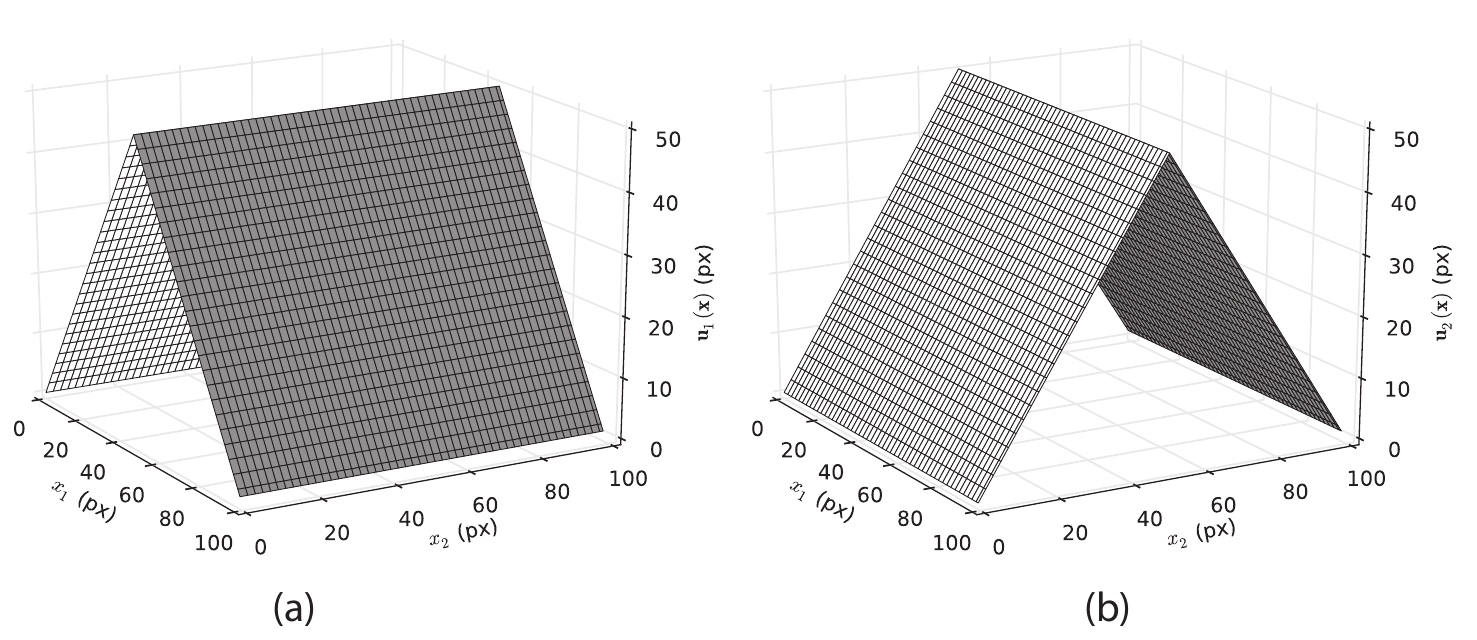}
  \caption{(a) Non-fully-differentiable displacement data $u_1(\mathbf{x})$ (b) $u_2(\mathbf{x})$}
    \label{fig:HatDisp}
\end{figure}

\begin{figure}[h!]
  \centering
    \includegraphics[scale=0.8]{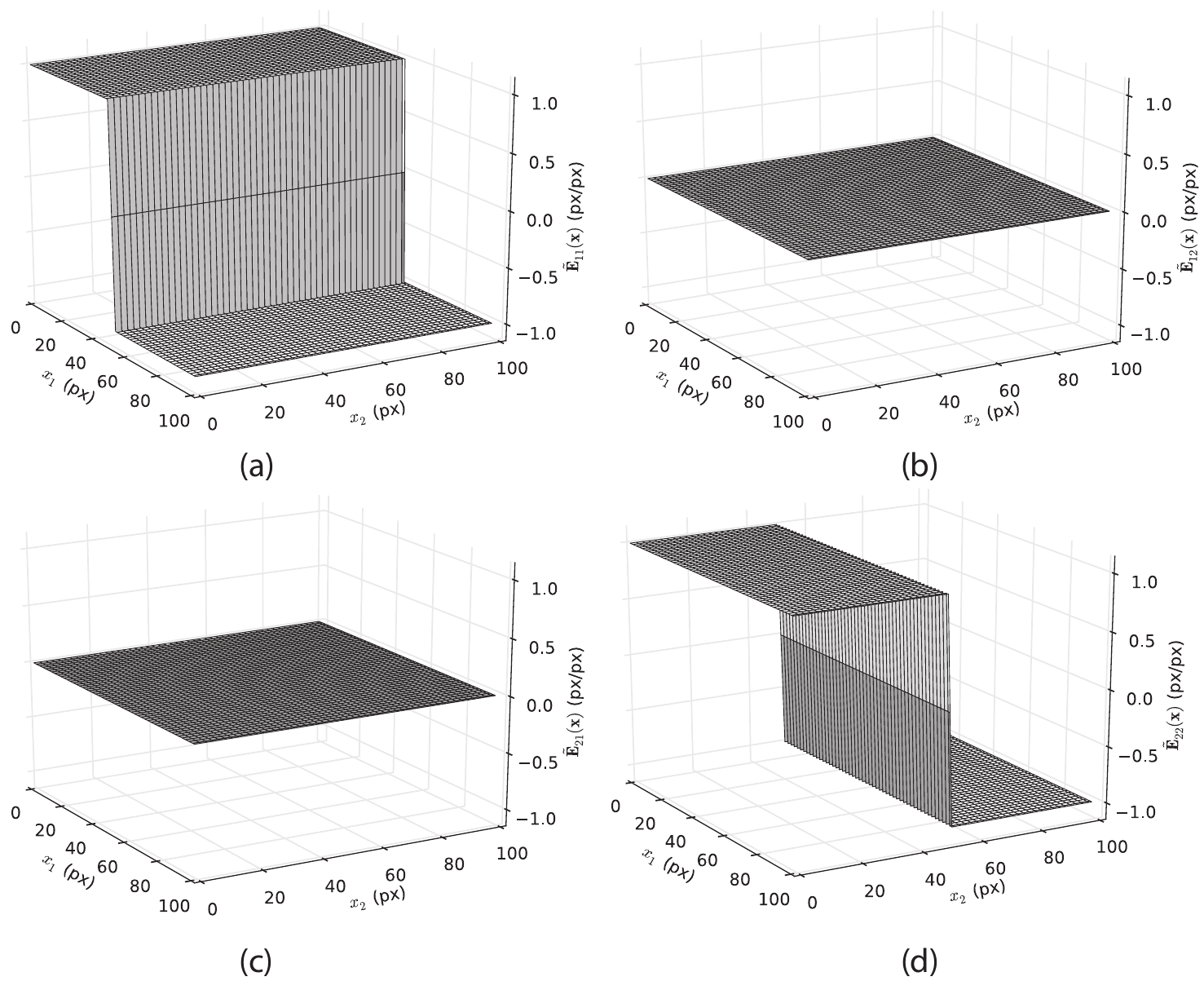}
  \caption{Components of the nonlocal strain tensor applied to the non-fully-differentiable displacement field (a) $\tilde{\mathbf{E}}_{11}$ (b) $\tilde{\mathbf{E}}_{12}$ (c) $\tilde{\mathbf{E}}_{21}$ (d) $\tilde{\mathbf{E}}_{22}$}
    \label{fig:HatStrain}
\end{figure}

The nonlocal strain, as detailed in \S\ref{sec:nl-str}, is well defined within the domain including at pixel interfaces and for fields that are continuous but not differentiable. To illustrate these two features, consider a square domain of size $L \times L$ units with a displacement profile of the form
\begin{equation}
\begin{aligned}
u_1(\mathbf{x}) &= \begin{cases}
x_1 & 0 < x_1 \leq L/2 \,\\
L - x_1 & L/2 < x_1 < L\, 
\end{cases} \\
u_2(\mathbf{x}) &= \begin{cases}
x_2 & 0 < x_2 \leq L/2 \,\\
L - x_2 & L/2 < x_2 < L\, 
\end{cases}
\end{aligned}
\end{equation}
Note that $\mathbf{u}(\mathbf{x})$ is not differentiable along the lines $x_1 = L/2$ and $x_2 = L/2$. The displacement profile is shown in Figure~\ref{fig:HatDisp}. The four components of the nonlocal strain tensor are shown in Figure~\ref{fig:HatStrain}. The nonlocal strain does not suffer from boundary effects if the support of $\boldsymbol{\alpha}_\epsilon$ is weighted properly towards the interior of the domain using the skew parameters $\gamma_1$ and $\gamma_2$. It is also important to point out that in this example the discontinuity lies on the interface between two subdomains $\Omega_j$. 
%

\subsection{Nonlocal strain and noisy data}

\begin{figure}[h!]
  \centering
    \includegraphics[scale=0.8]{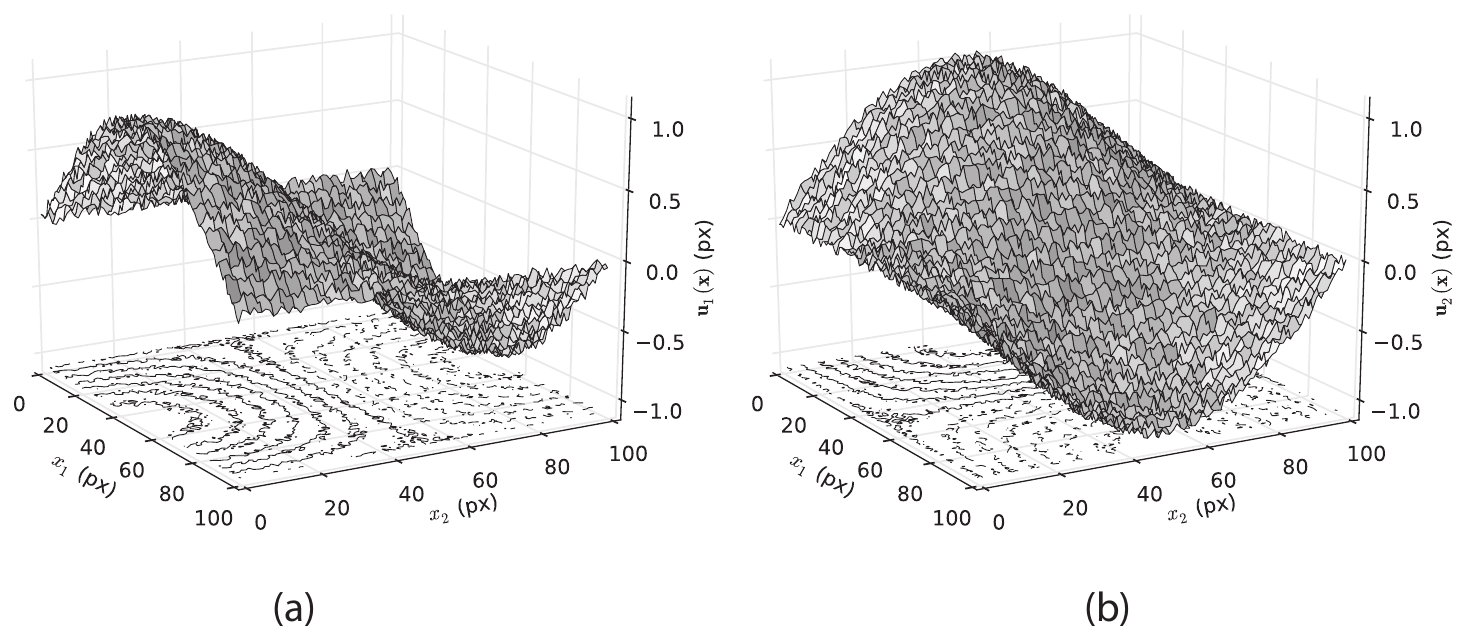}
  \caption{(a) Noisy displacement data $u_1(\mathbf{x})$ (b) $u_2(\mathbf{x})$}
    \label{fig:NoisyDisp}
\end{figure}

\begin{figure}[h!]
  \centering
    \includegraphics[scale=0.8]{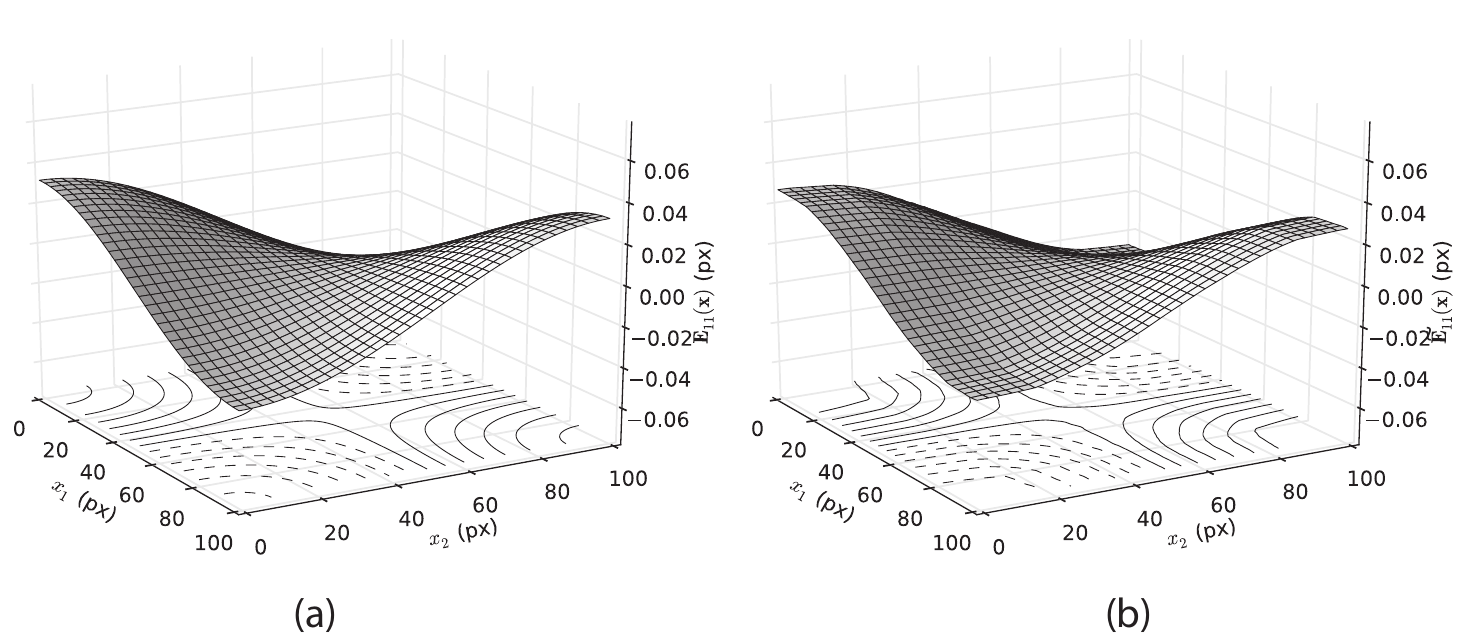}
  \caption{(a) Classical strain component $\mathbf{E}_{11}$ computed by taking the derivative of the displacement profile with no noise (b) nonlocal strain component $\tilde{\mathbf{E}}_{11}$ computed for the noisy displacement profile.}
    \label{fig:NoisyStrain}
\end{figure}

This numerical example demonstrates the effectiveness of the nonlocal strain measure for noisy data. The vector displacement profile for this example was generated by adding noise (with mean 0 and standard deviation 5\%) to a smooth function over a square domain of size 100 by 100 pixels. The displacement profile is given as
\begin{equation} \label{noisy-u}
\begin{aligned}
u_1(\mathbf{x}) &= \text{sin}\left( \frac{a}{2\pi} x\right)\text{cos}\left( \frac{a}{2\pi} y\right) + \sigma_{n1}  \\
u_2(\mathbf{x}) &= \text{cos}\left( \frac{a}{2\pi} x\right)\text{sin}\left( \frac{a}{2\pi} y\right) + \sigma_{n2} \; , 
\end{aligned}
\end{equation}
where $a=1/5$ and $\sigma_n$ is the amount of noise. Figure~\ref{fig:NoisyDisp} shows the noisy displacements plotted over the domain.
For the purpose of comparison, Figure~\ref{fig:NoisyStrain} shows the nonlocal strain component $\tilde{\mathbf{E}}_{11}$ computed for the noisy displacement data vs. the classical strain component $\mathbf{E}_{11}$ for the vector field \eqref{noisy-u} with the noise suppressed, i.e., $\sigma_{n1} = \sigma_{n2} = 0$. 

\begin{figure}[h!]
  \centering
    \includegraphics[scale=0.7]{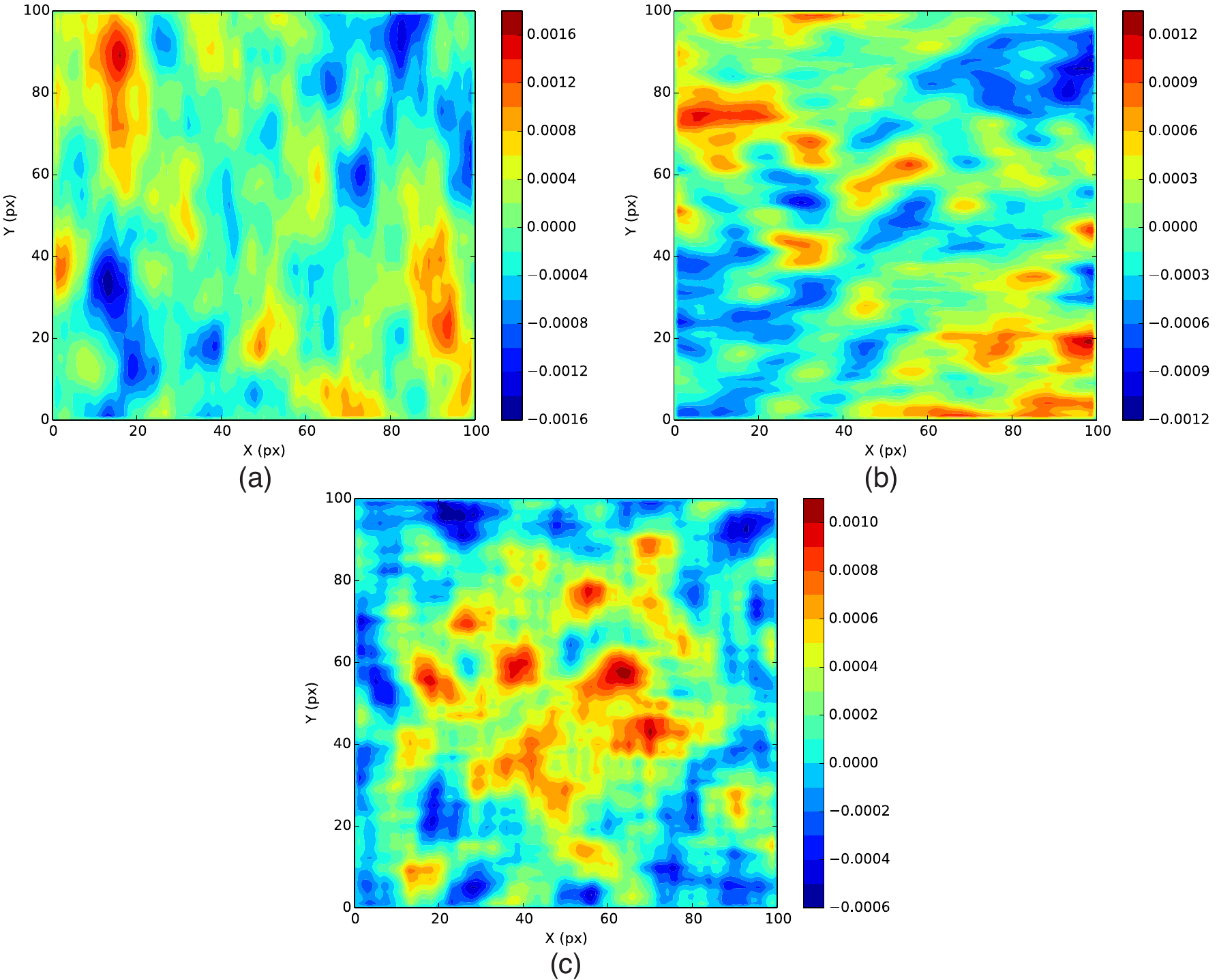}
  \caption{Error in the nonlocal strain components (a) $\tilde{\mathbf{E}}_{11}$ (b) $\tilde{\mathbf{E}}_{22}$ (c) $\tilde{\mathbf{E}}_{12}$.}
    \label{fig:NoisyStrainError}
\end{figure}

The nonlocal strain was computed using a support size, $\epsilon = 20$ units. This example demonstrates the robustness of the nonlocal strain measure in the context of noisy data and that the nonlocal strain is in agreement with the classical strain. Similar results hold for the other nonlocal strain components, although not shown here.
The error in each component of the nonlocal strain is shown in Figure~\ref{fig:NoisyStrainError}.

\subsection{DIC Challenge: synthetic strain concentrations of various periods and amplitudes}
This example shows the nonlocal strain measure as applied to one of the DIC Challenge \cite{DICChallenge} image sets with synthetic pixel displacements applied to induce strain concentrations of various periods and amplitudes. The data used in this example represent Sample 15 from the DIC Challenge images. The details of how the images were constructed are given on the DIC Challenge website. The reference image is \texttt{Ref.tif} and the deformed image is \texttt{P200\_K50.tif}. The correlation parameters used are given in Table \ref{tab:Sample15Params}. A step size of one pixel was used to enable evaluating the VSG and nonlocal strain measures with small support or small strain window size. It should be pointed out that the subset solutions are not independent at this step size \cite{Ke}.

\begin{figure}[h!]
  \centering
    \includegraphics[scale=0.4]{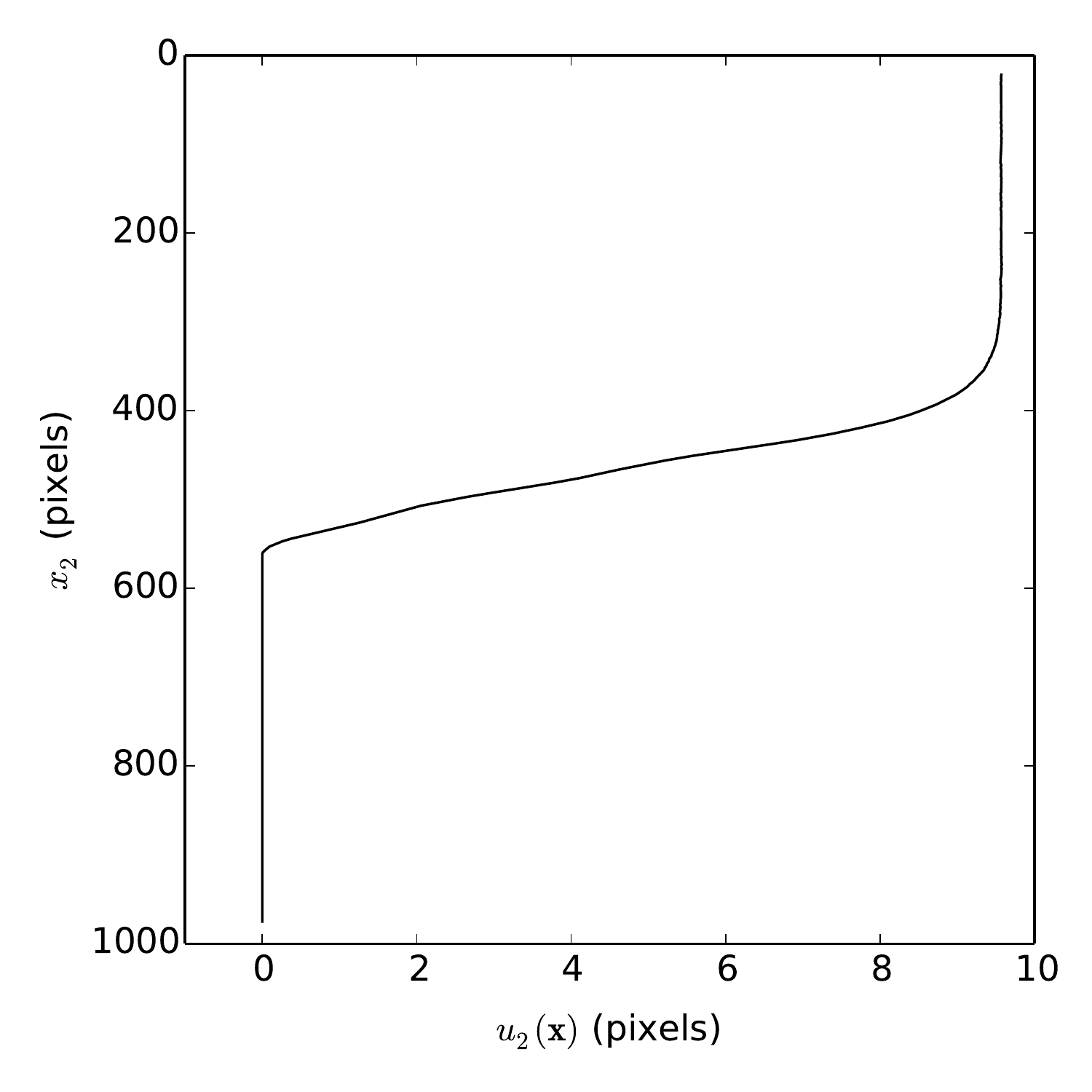}
  \caption{Displacement profile of the DIC Challenge, Sample 15 data set for a vertical line drawn through the images at $x = 1000$ pixels.}
    \label{fig:Sample15Disp}
\end{figure}

\begin{figure}[h!]
  \centering
    \includegraphics[scale=0.7]{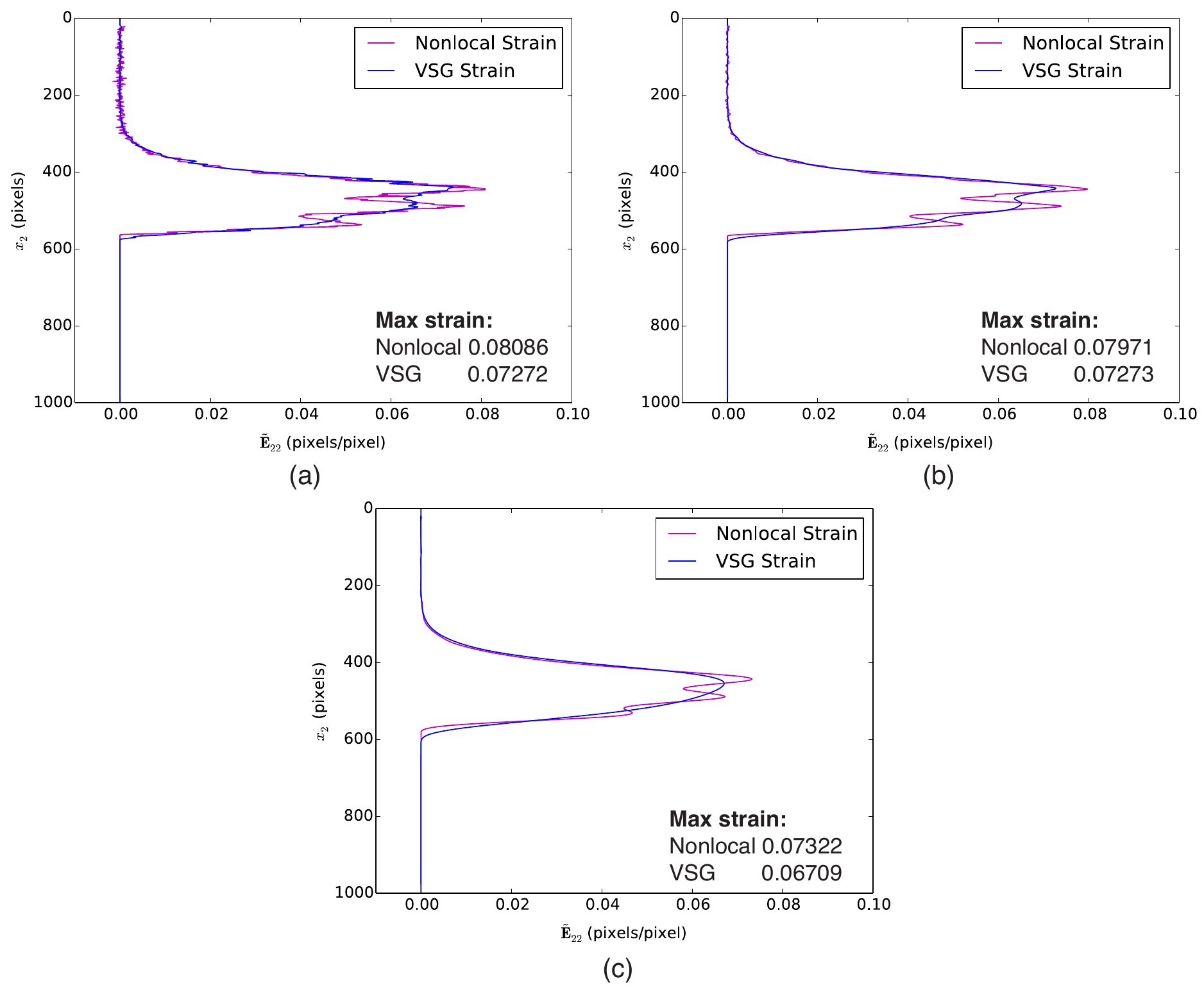}
  \caption{Nonlocal strain calculated for the DIC Challenge, Sample 15 images. (a) $\epsilon = 1$ pixel, strain window size = 1 (b) $\epsilon = 5$ pixels, strain window size = 5 (c) $\epsilon = 20$ pixels, strain window size = 20.}
    \label{fig:Sample15Strain}
\end{figure}

The displacement profile for a vertical line drawn through the images at $x = 1000$ pixels is shown in Figure~\ref{fig:Sample15Disp}. Figure~\ref{fig:Sample15Strain} shows the nonlocal strain results for this data set for various values of $\epsilon$ compared to the VSG method with various strain window sizes. To compute the VSG results, the Ncorr \cite{Ncorr} software was used. 

As expected, for larger values of $\epsilon$ the high frequency content of the nonlocal strain measure is effectively filtered out, but note the overall structure of the strain profile is preserved (multiple sub-peaks in the main strain concentration are captured). Using a larger strain window size also smooths the result, but the structure is not preserved. Reported in Figure~\ref{fig:Sample15Strain} is the maximum strain computed for each method, for each nonlocal support or strain window size. Taking the single pixel nonlocal support value of the maximum strain as the most accurate, the loss in accuracy going to a support size of 20 pixels is 9.4\%, whereas the loss in accuracy of the VSG method is 17.0\%, suggesting that smoothing the strain data with the VSG method leads to greater loss in accuracy than using the nonlocal strain.
 
\begin{table}[htb]
  \caption{Correlation parameters used for Sample 15 images\label{tab:Sample15Params}}
  \centering
  \begin{tabular}{l l}
    \hline
    Parameter & Value \\
    \hline
    Subset size & 25 \\
    Step size  & 1\\
    Interpolation & Bicubic \\
    Matching criterion & SSD \\
    Virtual strain gauge size & Varies from 1 to 20 \\
    Test function support, $\epsilon_1$ or $\epsilon_2$ & Varies from 1 to 20\\ \hline
  \end{tabular}
\end{table}

\subsection{DIC Challenge: experimental data for a plate with a hole being loaded in tension} This example involves computing the nonlocal strain from images taken of a steel plate with a small hole being loaded in tension. The images were obtained from the DIC Challenge website and represent Sample 12. Table \ref{tab:PlateHoleParams} lists the parameters used in the correlation to obtain the displacement values.
\begin{table}[htb]
  \caption{Correlation parameters for the plate with a hole images\label{tab:PlateHoleParams}}
  \centering
  \begin{tabular}{l l}
    \hline
    Parameter & Value \\
    \hline
    Subset size & 35 \\
    Step size  & 1 \\
    Interpolation & Bilinear \\
    Matching criterion & SSD \\
    Test function support, $\epsilon_1$ or $\epsilon_2$ & Varies from 5 to 50 pixels \\ \hline
  \end{tabular}
\end{table}
Note that a step size of one was chosen to introduce the maximum amount of high frequency content into the displacement solution. It is well known that increasing the step size leads to a smoother strain profile, but decreases accuracy as it introduces artificial dissipation of the displacement gradients. Using a virtual strain gauge approach to calculate the strain inevitably leads to a tradeoff between accuracy and smoothness. Large strain gradients cannot be captured if too much smoothing is introduced, leading to poor accuracy. Conversely, resolving steep gradients leads to a highly oscillatory solution. 

\begin{figure}[h!]
  \centering
    \includegraphics[scale=0.5]{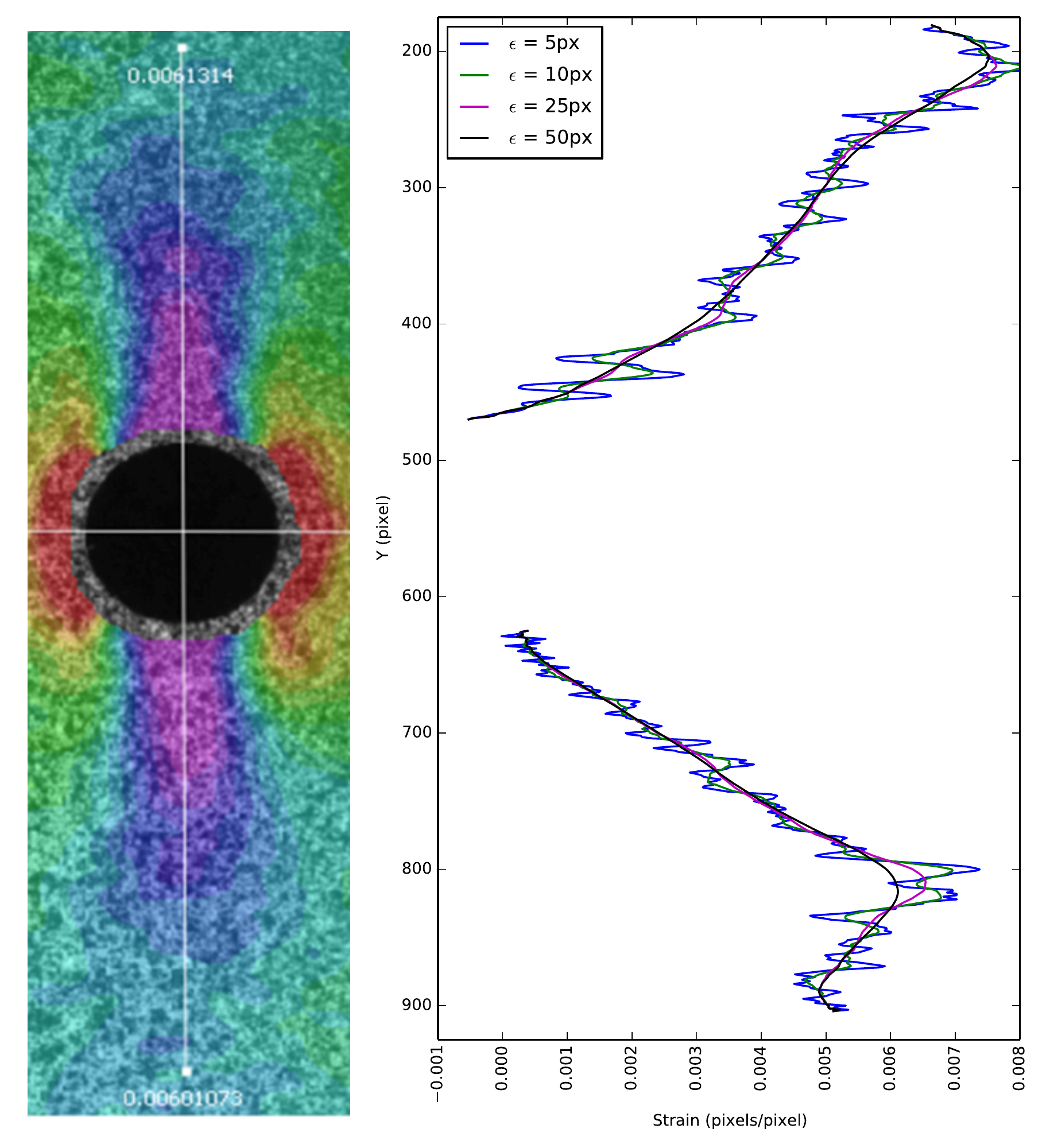}
  \caption{Plot of the nonlocal strain for a plate with a hole being loaded in tension for various values of $\epsilon$ (the width of the operator's support). The left image shows color contours of the principle strain as calculated using a virtual strain gauge in VIC 2D. Note that increasing $\epsilon$ has the effect of filtering high frequency noise in the displacement solution.}
    \label{fig:PHStrain}
\end{figure}

Figure \ref{fig:PHStrain} shows the strain calculated using equation \eqref{eq:NonlocalStrain} with the test function of equation \eqref{eq:HatFunction} for varying sizes of the test function support, $\epsilon$. It can be seen from Figure \ref{fig:PHStrain} that $\epsilon$ acts as a filter for high frequency content, but does not introduce dissipation, leading to a \emph{smooth and accurate strain profile}. 

\section{Conclusions}

We have introduced a new, nonlocal measure of strain for use in digital image correlation that is both noise filtering and appropriate for discontinuous displacement fields. With regard to noise, as opposed to curve fitting processes that discard data, the nonlocal process incorporates the full data set, but diminishes the effect of outliers by a distributed weighting of values that has a smoothing effect. In this way, the nonlocal strain maintains high fidelity data content without oscillations. Another feature of the nonlocal strain measure is that it gives the same result as the classical strain measure for differentiable displacement fields, but also provides a meaningful value when the displacement field is discontinuous (in which case the classical strain is not defined).  We have also shown that this strain measure is invariant under rigid body motion, which is necessary to prevent non-physical strains from arising due to large rotations or motion without deformation. Through a number of academic and experiment-based numerical examples we have demonstrated the effectiveness of this strain measure for both problems with analytic solutions and problems of engineering relevance. The results reveal that the nonlocal strain measure provides smoothing with less loss of accuracy than the VSG method. Ultimately, this work provides a framework to connect material characterization with experiments in a way that more fully incorporates data content and is robust enough to treat data with noise and discontinuities.

\section*{Acknowledgements}

This work was supported in part by Sandia National Laboratories. Sandia is a multiprogram laboratory 
operated by Sandia Corporation, a Lockheed Martin Company, for 
the United States Department of Energy's National Nuclear Security 
Administration under Contract DE-AC04-94AL85000.

This work was also supported in part by the Institute for Structural Engineering at Stellenbosch University. Their support is gratefully acknowledged.

\bibliographystyle{unsrt}
\bibliography{Master_References}

\end{document}